\documentclass{PoS}
\pdfoutput=1

\usepackage{subcaption}
\usepackage{wrapfig}
\usepackage{enumerate}
\usepackage{multirow}
\usepackage{booktabs}
\usepackage{bm}
\usepackage{array}
\usepackage{tabulary}
\usepackage{xspace}
\usepackage[normalem]{ulem}

\newcommand{\sv}{\langle \sigma \mathit{v} \rangle}
\newcommand{\svsvt}{\sv_\mathrm{svt}}
\newcommand{\svdsu}{\sv_{2.71}}
\newcommand{\hatsv}{\langle \widehat{\sigma \mathit{v}} \rangle}
\newcommand{\hatnu}{\hat{\bm{\nu}}}

\newcommand{\data}{\bm{\mathcal{D}}}
\newcommand{\lkl}{\mathcal{L}}

\newcommand{\mdm}{m_\mathrm{DM}}
\newcommand{\dom}{{\Delta\Omega}}

\newcommand{\bb}{b\bar{b}}

\newcommand{\tautau}{\tau^+\tau^-}

\newcommand{\Jobs}{J_\mathrm{obs,i}}
\newcommand{\lp}{\lambda_P}
\newcommand{\Junits}{GeV$^2$ cm$^{-5}$}
\newcommand{\svunits}{cm$^3$ s$^{-1}$}
\newcommand{\degree}{\ensuremath{{}^{\circ}}\xspace}

\title{Limits to dark matter properties from a combined analysis of
  MAGIC and {\it Fermi}-LAT observations of dwarf satellite galaxies}

\ShortTitle{MAGIC/{\it Fermi}-LAT combined dark matter searches}

\author{Javier Rico\\
  Institut de F\'{\i}sica d'Altes Energies (IFAE), E-08193 Bellaterra, Spain \\
  E-mail: \email{jrico@ifae.es}}

\author{\speaker{Matthew Wood}\\
  Kavli Institute for Particle Astrophysics and Cosmology,\\
  Department of Physics and SLAC National Accelerator Laboratory \\
  E-mail: \email{mdwood@slac.stanford.edu}}

\author{Alex Drlica-Wagner\\
  Center for Particle Astrophysics, Fermi National Accelerator Laboratory\\
  E-mail: \email{kadrlica@fnal.gov}}

\author{Jelena Aleksi\'c\\
  Institut de F\'{\i}sica d'Altes Energies (IFAE), E-08193 Bellaterra, Spain \\
  E-mail: \email{jelena@ifae.es}}

\author{for the MAGIC\footnote{http://magic.mpp.mpg.de} \, and
 the {\it Fermi}-LAT\footnote{http://www-glast.stanford.edu} \, Collaborations}

\abstract{We present the first MAGIC/{\it Fermi}-LAT joint search for dark
  matter annihilation gamma-ray signals from dwarf satellite
  galaxies. We combine 158 hours of observations of Segue~1 by MAGIC
  with 6-years observations of 15 dwarf satellite galaxies by the
  {\it Fermi}-LAT. We obtain limits on the annihilation cross-section for
  dark matter particle masses between 10 GeV and 100 TeV -- the widest
  mass range ever explored by a coherent and comprehensive
  analysis. Our new inclusive analysis approach is completely generic,
  and we propose to use it to perform a global, sensitivity-optimized
  dark matter search by combining data from present and future
  gamma-ray and neutrino detectors.}

\FullConference{The 34th International Cosmic Ray Conference,\\
		30 July- 6 August, 2015\\
		The Hague, The Netherlands}

\begin{document}
\section{Introduction}
\label{sec:intro}

Dark matter (DM) distributes in the Universe in halos that host galaxy
clusters, galaxies and galactic DM ``clumps''. A promising way to
identify the nature of DM and measure its properties is to search for
the Standard Model (SM) particles produced in its annihilation or
decay at these sites. Gamma rays and neutrinos are ideal messengers
for directional DM searches, since they are the only stable
neutral SM particles, and can thus travel from their production sites to
Earth unaffected by magnetic deflection.

Current gamma-ray instruments like the {\it Fermi}-LAT in space, the
ground-based Cherenkov telescopes MAGIC, VERITAS and H.E.S.S., and the
new-generation water Cherenkov detector HAWC, as well as neutrino
telescopes like IceCube and Antares, are sensitive to overlapping and
complementary DM particle mass ranges (from $\sim$1 GeV to $\sim$100
TeV).  All of these instruments have dedicated programs to look for DM
signals coming from, e.g., the Galactic center and halo
\cite{hessGH,magicGC,veritasGC,ref:FermiHalo,ref:hawcDM,ref:icecubeGH,ref:antaresDM}
, galaxy clusters
\cite{magicPerseus,hessFornax,veritasComa,ref:FermiClusters,ref:icecubeClusters},
or dwarf spheroidal satellite galaxies (dSphs) of the Milky Way (MW)
\cite{ref:MAGICSegue2014,veritasSegueErr,hessDwarfs,ref:Fermi2015}.

The universality of DM properties allows the combination of
data from different experiments and/or observational targets into a
global, sensitivity-optimized search \cite{ref:FullLikelihood}. For
a given DM particle model, a joint likelihood function can be
written as the product of the partial contributions from each of the
measurements/instruments. The advantage of such an approach is that
the details of each experiment do not need to be combined or averaged.
We have implemented this analysis framework, applicable to
observations from gamma-ray and neutrino instruments, and applied it
to the MAGIC and {\it Fermi}-LAT observations of dSphs.

The MW dSphs are associated to the Galactic DM sub-haloes,
predicted by N-body cosmological simulations, that have attracted
enough baryonic mass to start stellar activity (other sub-halos may
remain completely dark). MW dSphs have very high mass-to-light ratios,
being the most DM-dominated systems known so far
\cite{ref:Strigari}. MW dSphs also have the advantage of being free of
astrophysical gamma-ray sources and, while they are relatively close,
they still appear as quasi-point-like sources for gamma-ray and
neutrino telescopes, with relatively high expected fluxes. In
addition, the possibility of determining the DM distribution
and its uncertainty using a common methodology \cite{ref:Martinez}
allows a straightforward combination of their observations into a
sensitivity-optimized global analysis.

In this paper, we present our new global analysis framework and the
results of applying it to MAGIC and {\it Fermi}-LAT observations of dSphs.

\section{MAGIC and {\it Fermi}-LAT data samples}
\label{sec:instruments}
\label{sec:magic}

The \emph{Florian Goebel} MAGIC telescopes are located at the Roque de
los Muchachos Observatory (28.8$^\circ$~N, 17.9$^\circ$~W; 2200~m
above sea level), at the Canary Island of La Palma (Spain). MAGIC is a
system of two telescopes that detect Cherenkov light produced by the
atmospheric showers initiated by cosmic particles entering the Earth
atmosphere. Cherenkov images of the showers are projected by MAGIC
reflectors onto the photo-multiplier tube (PMT) cameras, and are used
to reconstruct the calorimetric and spatial properties of the primary
particle, as well as for its identification. Thanks to its large
reflectors (17 meter diameter), plus its high-quantum-efficiency and
low-noise PMTs, MAGIC achieves high sensitivity to Cherenkov light,
hence low energy threshold. The MAGIC telescopes are able to detect
cosmic gamma rays in the very-high-energy domain, i.e. in the range
between $\sim$50 GeV and $\sim$50 TeV.

For our study, we use MAGIC data corresponding to 158 hours of
observations of Segue~1 \cite{ref:MAGICSegue2014}, the deepest
observations of any dSph by any Cherenkov telescope.
The data were taken between 2011 and 2013 \cite{ref:MAGICupgrade2}.

The {\it Fermi}-LAT is a pair-conversion telescope sensitive
to gamma rays in the range from 20~MeV to more than 300~GeV
\cite{ref:Atwood2009}.  With its large field of view (2.4 sr), the
{\it Fermi}-LAT is able to efficiently survey the entire sky.  Indeed,
since its launch in August 2008, the {\it Fermi}-LAT has primarily
operated in a survey mode that scans the entire sky every
3 hours.  The survey-mode exposure coverage is fairly uniform over the
sky with variations of at most 30\% with respect to the average
exposure.  The {\it Fermi}-LAT source sensitivity, which is limited by
the intensity of diffuse backgrounds, shows larger variations but is
relatively constant at high galactic latitudes ($b > 10^\circ$).

In this work, we use a {\it Fermi}-LAT data sample corresponding to 6
years of observations of 15 dSphs, processed with the latest
(Pass 8) data analysis \cite{ref:Fermi2015}.  Events are selected with
energies between 500 MeV and 500 GeV in a $10\degree\times10\degree$
region of interest (ROI) centered on each dSph.  LAT likelihoods for a
given DM model are constructed from the bin-by-bin likelihoods of
\cite{ref:Fermi2015} and do not involve any reanalysis of the LAT
photon data.

\section{Analysis}
\label{sec:analysis}

\subsection{Dark Matter annihilation flux}

The gamma-ray (or neutrino) flux produced by DM annihilation in a
given target and detectable at Earth by an instrument observing a
field of view $\dom$ is given by:
\begin{equation}
\frac{d\Phi}{dE}(\dom) = \frac{1}{4\pi}\, \frac{\sv\, J(\dom)}{2\mdm^2}\,
\frac{dN}{dE}\quad ,
\label{eq:gammaflux}
\end{equation}
where $\mdm$ is the mass of the DM particle, $\sv$ the
thermally-averaged annihilation cross section, $dN/dE$ the average
gamma-ray spectrum per annihilation reaction (for neutrino this term
includes the oscillation probability between target and Earth), and
\begin{equation}
  J(\dom) = \int_{\dom}  d\Omega\ \int_\mathrm{l.o.s.} dl\,
  \rho^2(l,\Omega)
\label{eq:Jfactor}
\end{equation}
is the so-called \emph{astrophysical factor} (or simply J-factor),
with $\rho$ being the DM density, and the integrals running
over $\dom$ and the line of sight (l.o.s.), respectively.

Using PYTHIA simulation package version 8.205 \cite{ref:pythia}, we
have computed the average gamma-ray spectra per annihilation process
($dN/dE$) for a set of DM particles of masses between 10 GeV
and 100 TeV, annihilating into SM pairs $\bb$ and $\tautau$. For each
channel and mass, we average the gamma-ray spectrum resulting from
$10^7$ decay events of a generic resonance with mass $2\times\mdm$
into the specified pairs. For each simulated event, we trace all the
decay chains, including the muon radiative decay
($\mu^- \to e^- \bar{\nu}_e \nu_\mu \gamma$, not active in PYTHIA by
default), down to stable particles.

For the computation of the J-factors we follow the approach by
Martinez \cite{ref:Martinez}. The distribution of DM in the
halos of the different dSphs are parameterized following a
Navarro-Frenk-White profile (NFW) \cite{ref:nfw}:
\begin{equation} 
\rho(r) = \frac{\rho_0 r^3_s}{r(r_s+r)^2}\quad ,
\end{equation}
where $r_s$ and $\rho_0$ are the NFW scale radius and characteristic
density, respectively, and are determined from a fit to the dSph
stellar density and velocity dispersion profiles. The properties of
the dSphs used in our analysis, including the J-factors and their
uncertainties, can be found in \cite{ref:Fermi2015}.

We use templates for the DM emission in each dSph
normalized to its J-factor integrated to a radius of 0.5\degree from
the halo center ($\Jobs$).  The 0.5\degree integration region
encompasses more than 90\% of the annihilation flux for our dSph halo
models which have halo scale radii between 0.1\degree and 0.4\degree.
In the LAT analysis the intensity templates are used to construct a
three-dimensional model for the expected DM signal as a function of
space and energy within the $10\degree\times10\degree$ ROI centered on
each dSph.
The MAGIC analysis uses a one-dimesional likelihood for the photon
energy distribution within a signal aperture of radius
0.122$^\circ$. The observable flux in the MAGIC analysis of Segue~1 is
therefore reduced by a factor of $\sim$1.6 with respect to the LAT
analysis of the same target.  However we note that because the signal
aperture is matched to the angular size of the MAGIC PSF, this
truncation has a negligible impact on the sensitivity of the
analysis.

\subsection{Likelihood analysis}
\label{sec:lkl}

For each considered annihilation channel and DM particle
mass, we compute the profile likelihood ratio as a function of $\sv$:
\begin{equation}
\lp(\sv\, |\, \data) = \frac{\lkl(\sv; \hat{\hatnu}\,
  |\, \data)}{\lkl(\hatsv; \hatnu\,  |\, \data)}\quad ,
\label{eq:profile} 
\end{equation}
with $\data$ representing the data samples and $\bm{\nu}$ the
nuisance parameters. $\hatsv$ and $\hatnu$ are the values
maximizing the joint likelihood function ($\lkl$), and
$\hat{\hatnu}$ the value that maximizes $\lkl$ for a given
value of $\sv$. The likelihood function can be written as:
\begin{equation}
\lkl (\sv;\bm{\nu}\, |\, \data) =
\prod_{i=1}^{N_\mathrm{target}}  \lkl_i (\sv;  J_i,\bm{\mu}_i\, |\, 
\data_i) \cdot \mathcal{J}(J_i\, |\, J_{\mathrm{obs},i},\sigma_i)\quad
,
\label{eq:lkl-target}
\end{equation}
with the index $i$ running over the different targets (dSphs in our
case); $J_i$ is the J-factor for the corresponding target (see
Equation \ref{eq:Jfactor}); $\bm{\mu}_i$ denotes any additional
nuisance parameters; and $\data_i$ the target-related input data.
$\mathcal{J}$ is the likelihood for $J_i$, given measured
$\log_{10}(J_{\mathrm{obs},i})$ and its uncertainty $\sigma_i$
\cite{ref:Fermi2015}, i.e.:
\begin{equation}
  \mathcal{J}(J_i\, |\, \Jobs,\sigma_i) = \frac{1}{\ln(10) \Jobs \sqrt{2\pi}\sigma_i}
  \times
  e^{-\big(\log_{10}(J_i)-\log_{10}(\Jobs)\big)^2/2\sigma_i^2}\quad.
  \label{eq:jfactorPDF}
\end{equation}
The likelihood function for a particular target ($\lkl_i$) can in turn
be written as the product of the likelihoods for different
instruments (represented by the index $j$), i.e.:
\begin{equation}
  \lkl_i (\sv;  J_i,\bm{\mu}_i\, |\, \data_i) =
  \prod_{j=1}^{N_\mathrm{instrument}} \lkl_{ij}(\sv; J_i, 
  \bm{\mu}_{ij}\, |\, \data_{ij})\quad ,
\label{eq:lkl-instrument}
\end{equation}
where $\bm{\mu}_{ij}$ and $\data_{ij}$ represent the nuisance
parameters and input data sample for the given target $i$ and
instrument $j$.

Equations \ref{eq:profile}, \ref{eq:lkl-target} and
\ref{eq:lkl-instrument} are generic, i.e.\ they are valid for any set
of instruments and observed targets. In addition, they allow merging
of the results from different instruments and targets, starting from
tabulated values of $\lkl_{ij}$ vs.\ $\sv$ for a fixed value of $J_i$
and profiled with respect to $\bm{\mu}_{ij}$. These values can be
produced and shared by the different experiments without the need of
releasing or sharing any of the internal information used to produce
them.

For this work, we use the {\it Fermi}-LAT likelihood values vs.\ energy
flux, tabulated for different targets and in energy bins, and
released by the {\it Fermi}-LAT Collaboration \cite{ref:Fermi2015}. In the
case of MAGIC, the likelihood is obtained following the method
described in Refs.\ \cite{ref:FullLikelihood} and
\cite{ref:MAGICSegue2014}.

\section{Results}
\label{sec:results}

We compute one-sided, 95\% confidence level upper limits to $\sv$ by
numerically solving the equation $-2\ln \lp(\svdsu\, |\, \data) =
2.71$, for $\svdsu$. We consider $\lp$ restricted to non-negative
$\sv$ values only.

\begin{figure}[tbp]
\centering
\includegraphics[width=0.9\textwidth]{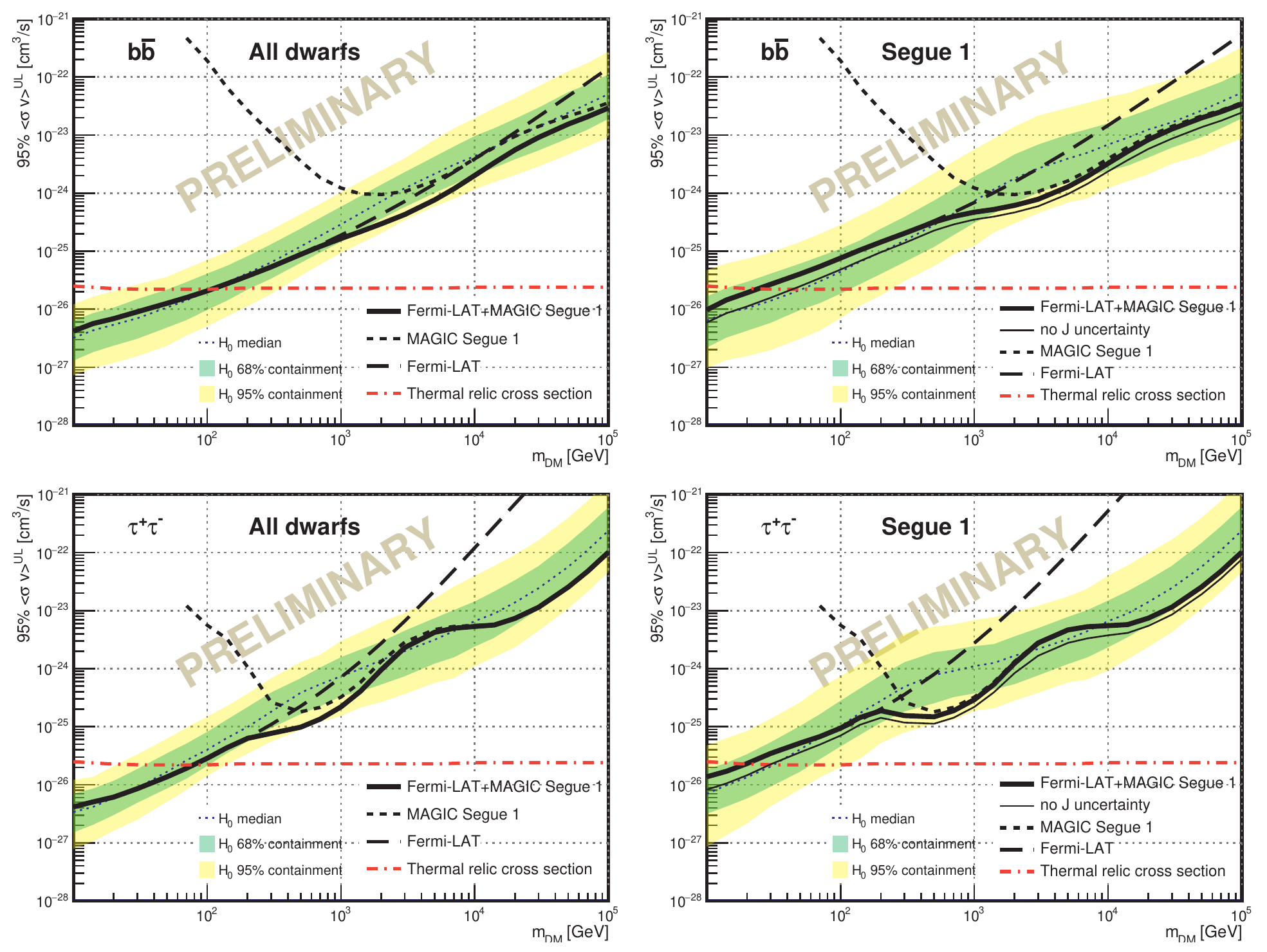}
\caption{95\% CL upper limits on the thermally-averaged cross section
  for DM particles annihilating into $\bb$ (upper plots) and $\tautau$
  (lower plots). Thick solid lines show the combined limits obtained
  by merging the {\it Fermi}-LAT observations of 15 dSphs (left plots)
  or Segue~1 (right plots) with MAGIC observations of Segue~1. Dashed
  lines show the individual MAGIC (short dashes) and {\it Fermi}-LAT
  (long dashes) limits.  J-factor statistical uncertainties are
  included.  For the Segue 1
  results, we also show (thin-solid line) the combined limits assuming
  a fixed J-factor (no statistical  uncertainties). The thin-dotted
  line, green and yellow bands show, 
  respectively, the median and the symmetrical, two-sided 68\% and
  95\% containment bands for the distribution of limits under the null
  ($H_0: \sv$=0) hypothesis (see main text for more details). The
  red-dashed-dotted line shows the thermal relic cross section from
  \cite{ref:Steigman2012}.}
\label{fig:alldwarfs} 
\end{figure}

Figure~\ref{fig:alldwarfs} shows the 95$\%$ confidence level limits to
$\sv$ for DM particles with masses between 10 GeV to 100 TeV
annihilating into SM pairs ($\bb$ and $\tautau$) obtained from the
combination of {\it Fermi}-LAT (15 dSphs and Segue~1 alone) and MAGIC
Segue~1 observations. The 68$\%$ and 95$\%$ containment bands are
computed from the distribution of upper limits obtained from 300
analysis realizations consisting of {\it Fermi}-LAT observations of
empty fields of view combined with MAGIC fast simulations (assuming in
both cases equal exposures as for the real data), and J-factors
randomly selected according to the probability density function (PDF)
in Equation~\ref{eq:jfactorPDF}.  The blank fields constituting the
LAT realizations were selected by choosing randomy sky positions with
$|b| > 30^\circ$ centered at least $0.5^\circ$ from a source in the
3FGL catalog.  MAGIC fast simulations consist of a set of event
energies randomly generated from the background PDF (see
Ref.~\cite{ref:MAGICSegue2014} for details) for both signal and
background regions.

We find no positive signal of DM in our data sample. As expected,
limits in the low and high ends of the considered mass range are
dominated by {\it Fermi}-LAT and MAGIC observations, respectively,
where the combined limits coincide with the individual ones. The
combination provides a significant improvement in the range between
$\sim$1 and $\sim$30 TeV (for $\bb$) or $\sim$0.2 and $\sim$2 TeV
(for $\tautau$), with a maximum improvement of the combined limits
with respect to the individual ones by a factor $\sim$2 at a mass of
500 GeV (for $\bb$) and 3 TeV (for $\tautau$).

MAGIC individual results shown here may differ from those presented in
Ref.\ \cite{ref:MAGICSegue2014} by up to a factor $\sim$4, which needs
a dedicated explanation. First, we note that the data, instrument
response functions, and likelihood functions are identical in both
works. Aside from enlarging the explored DM mass range, and in order
to homogenize MAGIC and {\it Fermi}-LAT analyses, we have introduced
the following differences between the two works: \emph{i)} the
J-factor for MAGIC cut of $J(0.122^\circ) = 2.2\times 10^{19}$ \Junits
(following Ref.\ \cite{ref:Martinez} and assuming an NFW DM density
profile); \emph{ii)} include the statistical uncertainties in the
determination of the J-factor; \emph{iii)} use the {\it Fermi}-LAT
prescription for limits close to bounds of the physical region
($\sv>=0$) (here we restrict the function $-2\ln \lp(\sv\, |\, \data)$
to $\sv \geq 0$, whereas in previous MAGIC results $\lp$ was also
computed for negative $\sv$ values, and the quoted limit was $\svsvt =
\svdsu-\hatsv$, whenever $\hatsv<0$, which is a more conservative
choice).

\section{Discussion and Conclusions}
\label{sec:discussion}

This work presents, for the first time, limits to the DM annihilation
cross-section from a comprehensive analysis of gamma-ray data of
energies between 500 MeV and 10 TeV. Using a common, homogeneous
analysis approach (both in the applied statistical methods and in the
determination of the J-factors), we have combined the MAGIC
observations of Segue~1 with {\it Fermi}-LAT observations of 15 dSphs.
This allowed the computation of meaningful global DM limits, and the
direct comparison of the individual results obtained with different
instruments. Our results span the DM particle mass range from 10 GeV
to 100 TeV -- the widest range covered by a single analysis to date.

We have not observed any DM signal. Consequently, we set
limits on the DM annihilation cross-section. Our results are the most
constraining from observations of dSphs in the considered mass range.
For the low-mass range, our results (fully dominated by {\it
  Fermi}-LAT data) are below the thermal relic cross-section
$\sv \simeq 3\times10^{-26} $ \svunits. In the intermediate mass range
(from few hundred GeV to few tens TeV, depending on the considered
annihilation channel), where {\it Fermi}-LAT and MAGIC achieve similar
sensitivities, the improvement of the combined result with respect to
the individual ones reaches a factor $\sim$2. In addition, we present,
for the first time, limits to high DM particle mass above 10 TeV
(fully dominated by MAGIC).

Our global analysis method is completely generic, and can be easily
extended to include data from more targets, instruments and/or
messenger particles provided they have similar sensitivity to the
considered DM particle mass range. Of particular interest is the case
of a global DM search from dSphs including data from all current
gamma-ray ({\it Fermi}-LAT, MAGIC, VERITAS, H.E.S.S, HAWC) and
neutrino (IceCube, Antares) instruments, and we hereby propose a
coordinated effort toward that end. Including results obtained from
other types of observational targets like the Galactic Center, galaxy
clusters or others is formally also possible, but a common approach to
the J-factor determination remains an open question. In the future,
this analysis could include new instruments like CTA, Gamma-400 or
Km3Net. Our global approach offers the best chances for indirect DM
discovery, or for setting the most stringent limits attainable by
these kinds of observations, therefore placing a new landmark in the
field.

\acknowledgments

The MAGIC Collaboration thanks the Instituto de Astrof\'{\i}sica de
Canarias for the excellent working conditions at the Observatorio del
Roque de los Muchachos in La Palma. We also acknowledge the financial
support of the ERDF under the Spanish MINECO (FPA2012-39502), of the
CPAN CSD2007-00042 and MultiDark CSD2009-00064 projects of the Spanish
Consolider-Ingenio 2010 program and of Centro de Excelencia Severo
Ochoa SEV-2012-0234.

The {\it Fermi}-LAT Collaboration acknowledges support for LAT
development, operation and data analysis from NASA and DOE (United
States), CEA/Irfu and IN2P3/CNRS (France), ASI and INFN (Italy), MEXT,
KEK, and JAXA (Japan), and the K.A.~Wallenberg Foundation, the Swedish
Research Council and the National Space Board (Sweden). Science
analysis support in the operations phase from INAF (Italy) and CNES
(France) is also gratefully acknowledged.

\end{document}